\documentclass[usenatbib,usedcolumn]{mn2e}
\usepackage{psfig,epsf}

\newbox\grsign \setbox\grsign=\hbox{$>$}
\newdimen\grdimen \grdimen=\ht\grsign
\newbox\laxbox \newbox\gaxbox
\setbox\gaxbox=\hbox{\raise.5ex\hbox{$>$}\llap
        {\lower.5ex\hbox{$\sim$}}}\ht1=\grdimen\dp1=0pt
\setbox\laxbox=\hbox{\raise.5ex\hbox{$<$}\llap
        {\lower.5ex\hbox{$\sim$}}}\ht2=\grdimen\dp2=0pt
\newcommand{\simlt}{\mathrel{\copy\laxbox}}

\newcommand{\Msolar}{\mbox{${\; {\rm M_{\sun}}}$}}
\newcommand{\oii}{\mbox{[O$\,${\sc ii}]}}
\newcommand{\oiii}{\mbox{[O$\,${\sc iii}]}}
\newcommand{\neiii}{\mbox{[Ne$\,${\sc iii}]}}
\newcommand{\nev}{\mbox{[Ne$\,${\sc v}]}}
\newcommand{\caii}{\mbox{Ca$\,${\sc ii}}}

\title[Radio galaxies in 2SLAQ]{Radio galaxies in the {2SLAQ}
  {L}uminous {R}ed {G}alaxy survey: {II}. The stellar populations of
  radio-loud and radio-quiet {LRGs}}
\author[H. M. Johnston et al.]{Helen M. Johnston,$^1$\thanks{E-mail:
  H.Johnston@physics.usyd.edu.au}, Elaine M. Sadler$^1$, Russell
  Cannon$^2$, Scott M. Croom$^1$, \newauthor 
  Nicholas P. Ross$^{3,4}$ and Donald P. Schneider$^4$\\ 
  $^1$School of Physics, University of Sydney, NSW 2006, Australia \\
  $^2$Anglo-Australian Observatory, PO Box 296, Epping, NSW 1710,
  Australia\\
  $^3$Department of Physics, University of Durham, South Road, Durham DH1 3LE\\
  $^4$Department of Astronomy and Astrophysics, The Pennsylvania State
  University, 525 Davey Laboratory, University Park, PA 16802}
\date{Accepted 2007 November 20.  Received 2007 November 14; in original
form 2007 September 19 }

\begin{document}

\maketitle

\label{firstpage}

\begin{abstract}
     
     We present an analysis of the optical spectra of a volume-limited
     sample of 375 radio galaxies at redshift {$0.4 < z < 0.7$} from
     the 2dF-SDSS Luminous Red Galaxy and QSO (2SLAQ) redshift survey.
     We investigate the evolution of the stellar populations and
     emission-line properties of these galaxies. By constructing
     composite spectra and comparing with a matched sample of
     radio-quiet sources from the same survey, we also investigate the
     effect on the galaxy of the presence of an active nucleus.
     
     The composite spectra, binned by redshift and radio luminosity,
     all require two components to describe them, which we interpret
     as an old and a younger population. We found no evolution with
     redshift of the age of the younger population in radio galaxies,
     nor were they different from the radio-quiet comparison sample.
     Similarly, there is no correlation with radio power, with the
     exception that the most powerful radio sources ($P_{1.4}>10^{26}
     \mathrm{W/Hz}$) have younger stars and stronger emission lines
     than the less powerful sources.  This suggests that we have
     located the threshold in radio power where strong emission lines
     ``switch on'', at radio powers of around
     $10^{26}\;\mathrm{W/Hz}$. Except for the very powerful radio
     galaxies, the presence of a currently-active radio AGN does not
     appear to be correlated with any change in the observed stellar
     population of a luminous red galaxy at $z\sim 0.5$.

\end{abstract}


\section{Introduction}
\label{sec:Introduction}

It is now accepted that most, if not all, massive galaxies host
super-massive black holes at their centre. The observed tight
correlation between the mass of the black hole and the global
properties of the host galaxy \citep{mtr+98,gkh+00} suggests that the
growth of the black holes at their centres is intimately related to
the formation and assembly of these galaxies.  However, at any one
time only $\sim$5\% of galaxies are seen as powerful radio sources or
show other evidence of an active galactic nucleus (AGN).  The question
arises: are the host galaxies of such AGN different from the hosts of
inactive sources?

It has been known since the early 1960s that the hosts for powerful
radio galaxies are massive ellipticals \citep[e.g.][]{mms64}.  The
trigger for the transition from quiescence to an active state has been
suggested to be gravitational interaction between galaxies.
Signatures of tidal interactions are common amongst radio galaxies
\citep[e.g.][]{hsb+86}, which presumably result in a large increase in
the amount of material feeding the black hole, thus triggering the
radio emission.  Some individual radio galaxies also show significant
association with young stellar populations
\citep[e.g.][]{att+01,tdm+02,wtrm02,jhcs05}; this suggests that the
merger process which triggers the radio emission can also lead to star
formation in the host galaxy. However, numerical studies of galaxy
interactions and mergers show that star formation is not necessarily
enhanced, due to the amount of gas swept out \citep{mcms07}.

Several questions remain to be answered in this scenario: What is the
relationship among mergers, the triggering of the active nucleus, and
star formation: are all three components necessarily present? Do all
early-type galaxies go through radio galaxy phases as they assemble?

In order to investigate these questions, we need to study a large and
uniform sample of radio galaxies.  Samples of this kind have recently
been studied in the local universe ($z<0.3$) by \citet{bkh+05} and
\citet{ms07}, but until now there has been no comparable sample
available at higher redshift.

\citet{bkh+05} investigated the properties of a sample of 2215 nearby
radio galaxies (most with 1.4\,GHz radio luminosity below
$10^{25}\,\mathrm{W/Hz}$) from the Sloan Digital Sky Survey.  They
showed that the fraction of galaxies hosting a radio-loud AGN was a
strong function of stellar mass, and that there was no correlation
between radio luminosity and optical emission-line luminosity for the
galaxies in their sample.  \citet{bkh+05} concluded that optical AGN
and low-luminosity radio-loud AGN are independent phenomena which are
triggered by different physical mechanisms.  \citet{ms07} studied a
sample of 2661 radio-loud AGN selected from the 6dF Galaxy Survey.
They confirmed the findings of \citet{bkh+05} that radio-loud AGN
preferentially inhabit the brightest and most massive host galaxies,
and showed that the fraction of all galaxies which host a radio-loud
AGN scales with the infrared K-band luminosity as $L_\mathrm{K}^2$.

We have constructed a volume-limited sample of radio sources by
combining optical data from the 2SLAQ LRG survey \citep{cde+06} with
the radio data from the VLA FIRST survey \citep{bwh95}. The 2SLAQ
sample is focused on luminous red galaxies, thus pre-selecting for
the large elliptical galaxies most likely to host powerful radio
sources. \citet{scm+07} used this sample to study the evolution of the
radio sources; in the current work, we investigate the properties of
the host galaxies.

The signal-to-noise ratio of individual 2SLAQ spectra is adequate for
redshift determinations, but is generally too low for accurate
measurement of weak spectral features.  We overcome this limitation by
combining spectra to form composites. This approach allows us to
investigate the variation of composites with both redshift and with
radio power, and to address the key question: How does the stellar
population of a galaxy correlate with the presence or absence of an
active radio-loud AGN in its centre?

Throughout this paper, we use $H_0$ = 71 km s$^{-1}$ Mpc$^{-1}$,
$\Omega_m$ = 0.27 and $\Omega_\Lambda$ = 0.73.

\section{Data reduction}
\label{sec:Data-reduct}

\subsection{Source selection}
\label{sec:source-selection}

The source selection is discussed in detail in
\citet[][Paper~I]{scm+07}; we present a brief summary here.

The 2SLAQ Luminous Red Galaxy survey \citep{cde+06} consists of 14,978
spectra taken with the Two-degree Field instrument (2dF) on the 3.9m
Anglo-Australian Telescope, of sources selected to have $ugriz$\ 
\citep{fig+96} photometry and non-stellar shape in the Sloan Digital
Sky Survey \citep[SDSS;][]{yaa+00}. The catalogue of sources with
spectra was then cross-matched with the VLA FIRST catalogue
\citep*{bwh95} and the NVSS \citep{ccg+98}.  These two surveys have
complementary properties, with NVSS sampling the total flux density of
extended sources, while FIRST has higher spatial resolution.  Galaxies
within 30\,arcsec of a FIRST/NVSS source were flagged as
possible matches (giving 2871 potential candidates).  Optical images
of the galaxies, primarily from Sloan Digital Sky Survey Data Release
3 \citep{aaa+05}, with a small number taken from the SuperCOSMOS Sky
Surveys \citep{hmr+01}, were compared visually with radio images from
the FIRST survey. This process yielded 391 optical galaxies with 2SLAQ
spectra which are true associations with radio sources. During this
visual inspection, we also flagged and accounted for sources which are
double and/or complex in structure.

Of these 391 radio sources, 14 had redshift quality values
$\mathrm{Q}<3$, indicating a less-reliable redshift determination.
These objects were excluded from further analysis; the remainder of
the objects, with $\mathrm{Q} \ge 3$, have redshift reliability $>
95\%$\ \citep[see][ \S 5.4]{cde+06}. The total radio detection rate
was 2.7\%. There were two sources in this list whose spectrum showed a
measurable redshift but was significantly contaminated by a foreground
star (the sources indicated as ``M star plus galaxy'' in Table~3 of
\citealt{scm+07}).  While these were usable for constructing a radio
luminosity function, they would complicate our spectral modelling, so
were removed from the sample for this paper, leaving 375 radio sources
with optical spectra.  

The total NVSS/FIRST radio flux of each object, including the flux
from all separate components identified by eye, was converted to total
radio power assuming $H_0 = 71\;\mathrm{km\,s^{-1}\,Mpc^{-1}}$\ and
$\Omega_\mathrm{M}=0.27$, using a $k$-correction of the form
$(1+z)^{-(1+\alpha)}$ with $\alpha=-0.7$, which is the median radio
spectral index \citep{mmb+03}. 

The final source list of 375 objects has redshifts ranging from
$z=0.31$\ to 0.76, and radio power between $P_\mathrm{1.4\,GHz} =
10^{23.44}\;\mathrm{W/Hz}$\ and $10^{27.02}\;\mathrm{W/Hz}$. 
The redshift distribution of radio sources is shown in
Fig.~\ref{fig:sourcedist}. The median redshift is $z=0.54$, and the
median radio power is $\log P=24.60$.


\subsection{Creation of a matched sample} 
\label{sec:matched}

To investigate the effect which the presence of an active nucleus has
on its host galaxy, we must create a sample of non-radio galaxies,
matched as closely as possible to the radio-loud sample.  Since the
probability that an early-type galaxy is a radio source is a strong
function of optical luminosity \citep{ape+77,sjk89}, the radio
galaxies in our sample are significantly more luminous than the 2SLAQ
LRG sample as a whole, which makes careful matching imperative.

For each radio source, we selected a matching source from the
remainder of the 2SLAQ catalogue. The sensitivity of the FIRST radio
catalogue \citep[with a typical flux density limit of
0.75~mJy;][]{bwh95} means that the upper limit to the radio power of
these ``radio-quiet'' sources range between $10^{23}$\ and
$10^{24}\;\mathrm{W/Hz}$\ at 1.4~GHz.  For each radio-loud object, we
found the object best matched in redshift and $r$-magnitude by
minimising the quantity
\begin{equation}
     \label{eq:match}
     d = (z_\mathrm{RL}-z_\mathrm{RQ})^2 +
     (r_\mathrm{RL}-r_\mathrm{RQ})^2/20 
\end{equation}

There are several possible ways of finding the ``nearest'' radio-quiet
galaxy; this expression was chosen empirically to weight the magnitude
and redshift equally, so the resulting sample has roughly the same
distribution in redshift and apparent magnitude.  Since radio-loud
sources make up only 2.7\% of the LRG sample, this approach produces a
sample which is well-matched to the radio-loud sources in redshift and
apparent $r$\ magnitude. The maximum discrepancy between a radio
source and its ``matched'' galaxy was 0.18~mag in apparent magnitude
and 0.0067 in $z$, with 90\% of sources having $\Delta r \simlt
0.03~\mathrm{mag}$ and $\Delta z \simlt 0.001$.

\subsection{Spectral reduction}
\label{sec:reduction}

We began with the wavelength-calibrated spectra produced by the 2SLAQ
project. These had a typical wavelength coverage of 5000--7250\AA,
which means that, at the redshift range of our targets, the region
containing \caii\ H and K and the 4000\AA\ break is always covered.
Although our spectra go well below 4000\AA\ rest wavelength, our
observational data are in the spectral region where instrumental and
atmospheric effects on continuum slope are not severe. For redshifts
beyond $z \sim 0.45$, H$\beta$\ and \oiii~5007 are shifted out of the
2dF wavelength range, so the most prominent emission line is
\oii~3727.

To combine the spectra, several steps of processing were applied to
each spectrum. The regions around the strongest night sky features
were removed and replaced by an interpolated value.  We are hampered
by the fact that the 2SLAQ spectra are not flux-calibrated, which
means that the instrument response can introduce changes in the slope
of the continuum. We attempted to correct for this, at least in a mean
sense, by dividing by an average response curve. We use the curve
derived by \citet{rpd+07}, which compared the 2SLAQ spectra of 160
galaxies which were also observed by the SDSS LRG survey (Eisenstein
et al. 2001, 2003)\nocite{eag+01,ehf+03}. We divided each spectrum by
this response curve: since we are producing composite spectra this
mean response curve should suffice, even though it does not take into
account night-to-night variations.  We then shifted each spectrum to
the rest frame using the known redshift, and rebinned all the spectra
onto the same scale, between 3280--4870\AA, with a dispersion of
1.4\AA/pixel

\subsection{Creation of composites}
\label{sec:composites}

\begin{figure}
     \centerline{\psfig{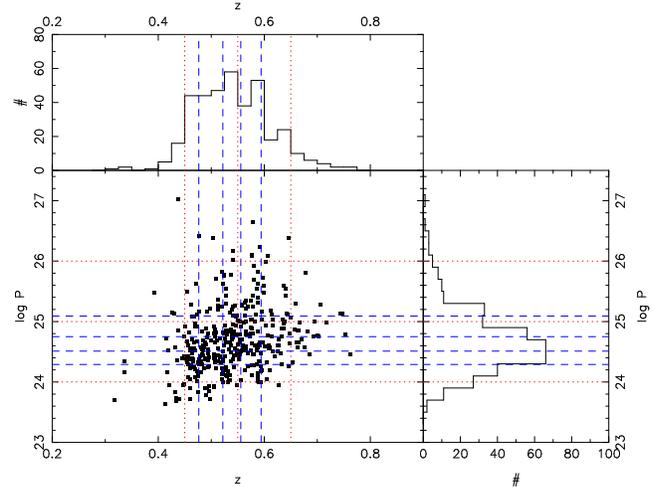}}
     \caption{Distribution of sources in redshift and radio power at
       1.4~GHz (measured in $\mathrm{W/Hz}$). The separate histograms
       of redshift and radio power are shown above and to the right,
       respectively. The binning for the four separate sets of
       composites are indicated by the lines. Set A, binning by
       redshift with 75 sources in each bin, is shown by the vertical
       blue dashed lines; set B, binning by redshift with equal
       spacing, is shown by the vertical red dotted lines. Set C,
       binning by radio power with 75 sources in each bin, is shown by
       the horizontal blue dashed lines; set D, with equally spaced
       bins in radio power, is shown by the horizontal red dotted
       lines.}
     \label{fig:sourcedist}
\end{figure}

The distribution of sources in ($z$,$\log P$) is shown in
Fig.~\ref{fig:sourcedist}.  Although the redshifts range from $z=0.31$
to $0.76$, and the radio power from $\log P=23.44$ to $27.02$, the
bulk of sources cluster near the middle of the range in each
parameter. We thus had two choices: we can create composites with
equal numbers of sources, which produces composites with equal
signal-to-noise, or we can create composites within equal intervals in
redshift or radio power, which maximises the baseline over which we
can investigate changes, but at the expense of lowering the
signal-to-noise ratio of the outlying composites.

We thus created \emph{four} separate sets of composite spectra
(Table~\ref{tab:composites}, shown graphically by the dashed lines in
Fig.~\ref{fig:sourcedist}):
\begin{enumerate}
\item[(A)] sort spectra by redshift and create 5 composites with
  equal numbers of sources: each composite is composed of 75 sources,
  and the redshift limits are shown in Table~\ref{tab:composites};
  
\item[(B)] sort by redshift and create 4 composites with equal spacing
  in redshift: this gave us four groups, with $z<0.45$ (28 sources),
  $0.45<z<0.55$ (196 sources), $0.55<z<0.65$ (138 sources), and
  $z>0.65$ (28 sources);

\item[(C)] sort by radio power and create 5 composites with
  equal numbers of sources: each composite is composed of 75 sources,
  and the redshift limits are shown in Table~\ref{tab:composites};

\item[(D)] sort by radio power and create 4 composites with equal
  spacing in radio power: this gives us four bins, with $\log P < 24$
  (27 sources), $24 < \log P < 25$ (269 sources), $25 < \log P < 26$
  (83 sources), and $\log P > 26$ (11 sources). 

\end{enumerate}

\begin{table*}
 \caption{Parameters of the composites and results of the fits. Four
   sets of composites A--D were created (see \S~\ref{sec:composites}
   for details). Columns 1--4 show the name of the composite set, the
   range in redshift or radio power, and the number of spectra which
   went into each composite. The next three columns show the results
   of the spectra modelling (\S~\ref{sec:modelling}): the age of the
   young population, the fraction of light contributed by the young
   population at 4050\AA, and the reduced $\chi_\nu^2$ value
   for the best fit (for $\nu=31$ degrees of
   freedom). Corresponding composites of a matched sample of
   radio-quiet sources (\S~\ref{sec:matched}) were created, and the
   results of fitting to these are shown in columns 8--10.
  }
\label{tab:composites}
\begin{tabular}{lllr rcc rcc}
\hline
       &    &       &        & 
  \multicolumn{3}{c}{Radio-loud sample} &
  \multicolumn{3}{c}{Radio-quiet sample} \\
       & &    & No. of & Age   &  & & Age   &  & \\
Sample & \# & Range & galaxies & 
  (Myr) & Fraction & $\chi_\nu^2$ &
  (Myr) & Fraction & $\chi_\nu^2$ \\
\hline
              & 1 & $z=0.31$--$0.48$ & 75 &  690 & 0.40 & 3.55 &  710 & 0.54 & 4.42 \\
(A) Redshift: & 2 & $z=0.48$--$0.52$ & 75 & 1000 & 0.47 & 3.61 & 1130 & 0.56 & 2.48 \\
equal numbers & 3 & $z=0.52$--$0.56$ & 75 &  700 & 0.45 & 3.29 &  690 & 0.46 & 1.84 \\
              & 4 & $z=0.56$--$0.59$ & 75 & 1120 & 0.55 & 1.32 &  670 & 0.50 & 0.87 \\
              & 5 & $z=0.59$--$0.76$ & 75 & 1130 & 0.60 & 1.32 & 1120 & 0.70 & 1.55 \\
\\
                & 1 & $z=0.31$--$0.45$ & 26  &  690 & 0.35 & 6.48 & 1110 & 0.57 & 2.00  \\
(B) Redshift:   & 2 & $z=0.45$--$0.55$ & 192 &  700 & 0.40 & 3.26 & 1000 & 0.54 & 3.19  \\
equal spacing   & 3 & $z=0.55$--$0.65$ & 133 & 1120 & 0.66 & 1.26 & 1120 & 0.66 & 1.52  \\
                & 4 & $z=0.65$--$0.76$ & 24  & 1130 & 0.64 & 0.94 & 1130 & 0.81 & 0.55  \\
\\
                 & 1 & $\log P=23.44$--$24.29$ & 75 &  720 & 0.50 & 3.61 &  710 & 0.53 & 2.68 \\
(C) Radio power: & 2 & $\log P=24.29$--$24.51$ & 75 & 1120 & 0.52 & 1.90 & 1010 & 0.59 & 1.42 \\
equal numbers    & 3 & $\log P=24.51$--$24.75$ & 75 &  710 & 0.45 & 1.94 &  710 & 0.64 & 6.45 \\
                 & 4 & $\log P=24.75$--$25.09$ & 75 &  700 & 0.47 & 2.06 & 1130 & 0.58 & 0.90 \\
                 & 5 & $\log P=25.09$--$27.02$ & 75 & 1140 & 0.53 & 1.06 &  700 & 0.56 & 1.90 \\
\\
                 & 1 & $\log P=23.44$--$24$ & 26  &  710  & 0.55 & 3.81 &  710 & 0.45 & 3.00 \\
(D) Radio power: & 2 & $\log P=24$--$25$    & 259 &  700  & 0.46 & 2.84 &  710 & 0.55 & 2.84 \\
equal spacing    & 3 & $\log P=25$--$26$    & 79  & 1130  & 0.51 & 0.81 &  990 & 0.64 & 1.77 \\
                 & 4 & $\log P=26$--$27.02$ & 11  &  110  & 0.46 & 1.61 &  680 & 0.59 & 0.68 \\
\hline
\end{tabular}
\end{table*}

There are several different approaches which may be taken in combining
low signal-to-noise ratio spectra to form composites: spectra may be
weighted by the variance in each spectrum
\citep[e.g.][]{fhf+91,ehf+03}, or weighted by luminosity
\citep{bgb+02,dop+04}; or the composite can be constructed from the
median of individual spectra at each wavelength \citep{crc+02}. Since
the 2SLAQ galaxies are all intrinsically luminous red galaxies, we
have taken the simplest approach and created composites by averaging
the spectra in each group.  The spectra were all taken with very
similar wavelength coverage but have different redshifts, so in each
rest wavelength bin, different numbers of spectra contribute to the
sum, particularly for the highest and lowest redshifts. We corrected
for this by counting the number of spectra which contribute at each
wavelength and dividing the composite by this number.

The composite spectra were then normalised by the flux at 4050\AA.
Since the resolution of the spectral models we chose is 10\AA\ (see
\S~\ref{sec:modelling}), we binned each composite into 20\AA\ bins,
and then removed bins which included the bright emission lines,
principally \oii\ and \oiii. We estimated the errors on the flux in
each wavelength bin by measuring the RMS of the 14 original pixels
which were included in each 20\AA\ bin. 
We chose to perform the fits to the models over the wavelength range
3500--4400\AA, based on our desire to fit the same wavelength region
in all spectra; a total of 33 independent data points in each spectrum were used in the fits.

Composites for the matched sample of radio-quiet sources
(\S~\ref{sec:matched}) were created in exactly the same manner; this
data set serves as a control sample to see how the stellar population
varies in the absence of a powerful radio source.

\section{Data analysis}
\label{sec:analysis}

\subsection{Differences between composites}
\label{sec:differences}

To determine if the presence of an active nucleus affects the galaxy,
we compared the composites formed from the radio-loud sample and the
matched radio-quiet sources (\S~\ref{sec:matched}). The most basic
test is to use a simple $\chi^2$ test to find the probability that a
pair of composites come from the same underlying distribution.  Each
of the eighteen radio-loud composite spectra
(Table~\ref{tab:composites}) was compared with the corresponding
radio-quiet counterpart. All composites were consistent with being
from the same distribution, except the D4 composite formed from the
highest-power radio sources, where the $\chi_\nu^2=3.5$. This initial
evidence suggests that, with the exception of the very highest-power
radio sources, there is no evidence of difference between galaxies
containing an active AGN and those that do not.

\subsection{Modelling the stellar continuum}
\label{sec:modelling}

We modelled the observed continuum as an old population plus a blue
component due to a second population of stars of a younger age, using
models from the \textsc{gissel96} library \citep{bc93}. Of the models
in their library, we used the ones with a Salpeter initial mass
function with mass limits of 0.1 and 100\Msolar\ and solar
metallicity, using the \citet{gs83} stellar spectral atlas. 

We used two single-age stellar populations, representing instantaneous
bursts of star formation. The old population was represented by a
single-age population of age 7000~Myr, the younger stars by a
population with a range of ages (10, 20, 50, 70, 100, 200, 500, 700,
1000, 2000 or 5000~Myr). The exact age of the ``old'' population is
not critical, since the spectrum changes little at these ages; we
chose 7~Gyr since at the highest redshift in our sample ($z=0.76$) the
age of the universe was 7~Gyr. We assembled the model spectra from the
\textsc{gissel96} library, and binned them to the same resolution as
the observed spectra.

Since the observed continuum is modelled as the sum of two
populations, we assumed each composite is represented as a 7000~Myr
population plus some fraction $f$ of a younger population with age
$\tau$; our task is to determine $f$ and $\tau$ for each composite.
This parametrisation is unlikely to be a realistic representation of
the stellar populations of these galaxies. However, the exact form of
the models is not important, because our study is a
\emph{differential} test, comparing the properties of the radio-loud
galaxies with their radio-quiet counterparts.

We created a $11 \times 11$\ grid of model spectra. Along one axis, the
age of the young population $\tau$\ is varied between 10~Myr and
5~Gyr; along the other axis the fraction of light contributed by the
young population varies linearly between 0 and 100\%. Thus the
spectrum at grid position (5,3) consists of an old (7 Gyr) population
plus a 100~Myr population, where the young population is contributing
20\% of the light at 4050\AA.

We calculated the $\chi^2$ difference between the model spectrum and
each of the observed composite spectra for all 121 model spectra. We
allowed the normalisation of the model spectrum to vary between 0.75
and 1.25 of the flux of the observed spectrum at 4050\AA, to allow for
possible differences in slope between the model and the observed
spectrum. This yields a $\chi^2$ map for each observed composite; the
best fit is chosen to be that where the $\chi^2$ value is minimised.

\begin{figure}
     \centerline{\psfig{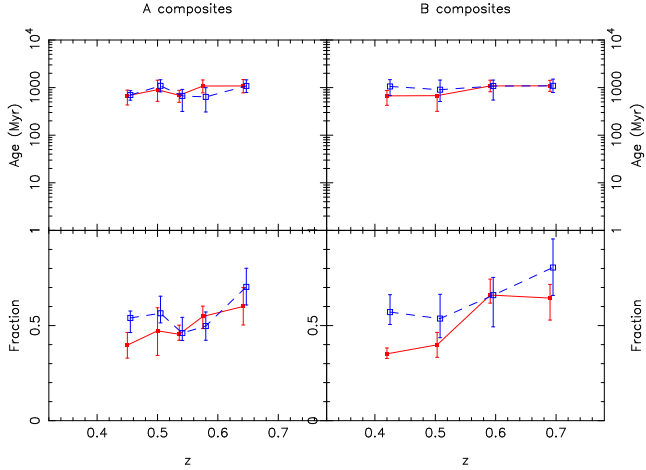}}
     \caption{Results of fits to the A and B composites, showing the
       age and fraction of light contributed by young stars as a
       function of redshift. The red filled symbols (solid lines)
       represent the radio-loud sources, while the blue open symbols
       (dashed lines) represent the matched sample of radio-quiet
       sources; the radio-quiet sources have been offset slightly in
       redshift for clarity. The age of the young population does not
       change with redshift; the fraction of light contributed rises
       slightly.}
     \label{fig:zfits}
\end{figure}

\begin{figure}
     \centerline{\psfig{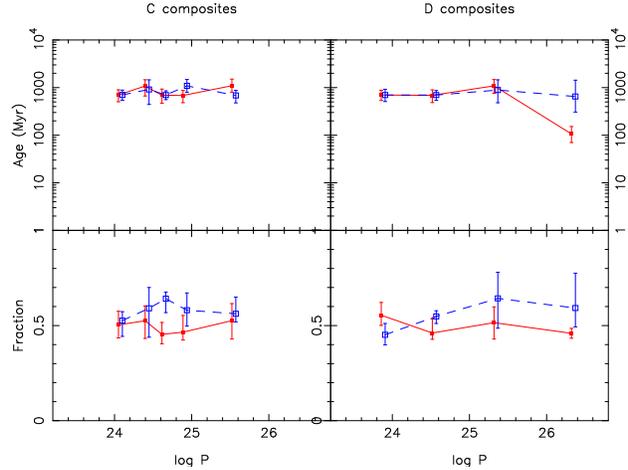}}
     \caption{Same as Fig.~\ref{fig:zfits}, but for the C and D
       composites, showing the age and fraction of light contributed
       by young stars as a function of radio power. Neither the age of
       the young population nor the fraction of light contributed
       changes significantly as a function of radio power,
       \emph{except} at the highest radio power (composite D4), where
       the second population is significantly younger. }
     \label{fig:Pfits}
\end{figure}

The results of the fitting are shown in Table~\ref{tab:composites},
and shown graphically in Figures~\ref{fig:zfits} and \ref{fig:Pfits}.
The best-fitting age and fraction were derived from a parabolic fit to
the one-dimensional $\chi^2$ curves through the minimum. The error
bars represent the confidence interval given by the ellipse
representing $\Delta\chi^2=3$\ (containing $\sim90\%$ of normally
distributed data). In all cases, the $\chi^2$ of the two-component
model fit was significantly ($\sim$ factor 2) better than the $\chi^2$
of the fit using a single-aged stellar population.

The results of our modelling can be summarised as follows. There is no
evolution with redshift of the stellar population of the radio
galaxies, nor are they in any way different from radio-quiet galaxies
matched in redshift and $r$ luminosity. The age of the ``young''
population is unchanged with redshift, with our spectra requiring a
second population of stars of age $\sim 820\;\mathrm{Myr}$ over the
whole redshift range of our sample. The fraction of light contributed
by young stars is about 50\%, and while it rises slightly over the
redshift range of the sample, the change is not significant ($1.4\sigma$). 

\begin{figure}
     \centerline{\psfig{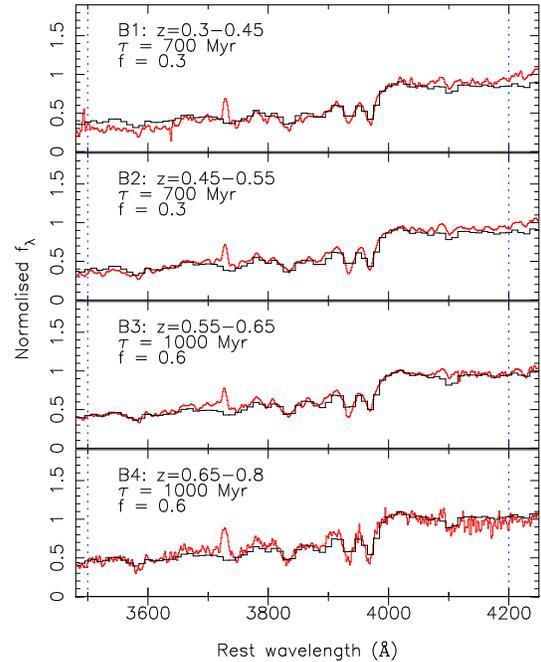}}
     \caption{Model fits to the four B composite spectra, with equal
       spacing in redshift. The composite spectrum is shown in red,
       with the model fit, consisting of an old population plus a
       young population of age $\tau$\ contributing a fraction $f$\ of
       the light at 4050\AA, shown as the thick black line
       (Table~\ref{tab:composites}). The dotted lines indicate the
       wavelength range over which the fitting was performed. }
     \label{fig:modelB}
\end{figure}

The B-composites and their best-fitting models are shown in
Fig.~\ref{fig:modelB}. 

\begin{figure}
     \centerline{\psfig{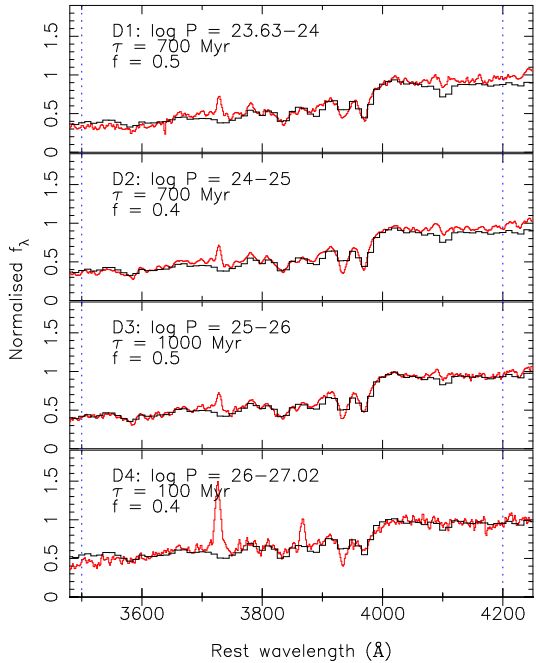}}
     \caption{Model fits to the four D composite spectra, with equal spacing
       in radio power. The composite spectrum is shown in red, with
       the model fit shown as the thick black line
       (Table~\ref{tab:composites}). Note the substantially stronger
       \oii\ emission line in the highest radio power spectrum
       (composite D4), as well as \neiii\ emission.}
     \label{fig:modelD}
\end{figure}

We can also examine how the spectrum changes as a function of radio
power, using the C and D composites defined earlier. Again, our
modelling shows no change in the age of the population or the fraction
of young stars as the radio power increases (Fig.~\ref{fig:Pfits}).
The only exception is that the very highest power sources (the D4
composite, consisting of the 11 sources with $\log P > 26$;
Table~\ref{tab:composites}) appear to have a substantially younger
population (100~Myr), as well as strong \oii\ 3727 emission (which was
not included in the fit).  This difference is not seen in the matching
sample of radio-quiet sources, suggesting that the young stars are in
some way associated with the active nucleus.  Recall that this was the
only composite showing a simple $\chi^2$ difference with the other
composites (\S~\ref{sec:differences}). We checked the eleven
individual spectra which went into the D4 composite, in case the
composite was being skewed by a single peculiar object. No individual
spectrum appeared to be peculiar; 6 of the 11 showed strong emission
lines. The D-composites and their best-fitting models are shown in
Fig.~\ref{fig:modelD}. We tried splitting the second-highest power bin
(D3), to see if there was a continuous trend; both sub-bins had
similar ages, of $\sim 1000$ and $\sim 700$~Myr, quite different from
the 100~Myr population found in the D4 bin. Thus the difference
appears to be confined to the very highest radio-power sources.

\begin{figure}
     \centerline{\psfig{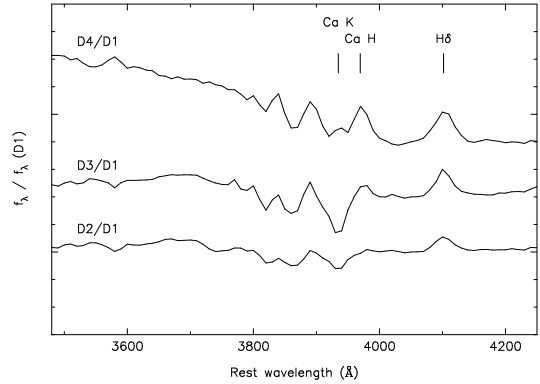}}
     \caption{Comparison between the model fits to the four D
     composite spectra (Fig.~\protect\ref{fig:modelD-comp}). The
     models are shown divided by the best-fit model to the D1
     composite ($\tau=500$~Myr, $f=0.5$), and offset in the vertical
     direction for clarity.}
     \label{fig:modelD-comp}
\end{figure}

In Figure~\ref{fig:modelD-comp} we plot the ratio of these models. The
figure shows there are small changes to spectral lines, but the
principal difference is a change in slope in the model fit to the D4
composite.

We investigated whether the young population in the highest-power
radio sources could in fact be due to emission from the central AGN,
which would also produce excess blue light. We tried replacing the
second, young population in our fits with a power-law continuum of the
form $f_\lambda \propto \lambda^\alpha$, where $\alpha$\ was allowed
to vary between $-5$\ and 0. The fit was significantly poorer than the
best-fitting model with a young stellar population: the 4000\AA\ break
was not well reproduced. Thus we conclude that the highest power radio
sources do show evidence of association with a population of young
stars.

\subsection{Emission lines}
\label{sec:emission-lines}

Emission lines were present in many of the spectra, mostly \oii, and
occasionally \neiii\ and \nev\ \citep[see Table~3 of][]{scm+07}. The
\oiii\ 4958,5007 pair was seen in only a single galaxy,
J100322.41-000137.8, but was redshifted beyond the wavelength range of
most of our sources.  \citet{scm+07} found that the fraction of 2SLAQ
radio galaxies which show \oii\ emission is higher (27\%) than the
overall 2SLAQ spectroscopic sample \citep[17.7\%;][]{rpd+06}. However,
since more luminous galaxies are more likely to show emission, this
result needs to be approached with caution. In any case, most of the
radio galaxies in 2SLAQ would not have been recognised as AGN on the
basis of their optical spectra alone. This agrees well with several
other studies, which found little evidence of correlation between
radio power and emission line strength in nearby radio galaxies
\citep{rwb91,olk95}.

By using a comparison sample of radio-quiet galaxies, matched in
redshift and luminosity, we are able to investigate this
effect. Comparison of the emission-line properties of our radio-loud
sources with the matched sample of radio-quiet galaxies reveals
differences between the two groups. We can examine the emission lines
in two ways: finding the number of individual galaxies which show
emission lines as a function of various parameters, and investigating
the properties of the emission lines in the composite spectra.
 
\subsubsection{Emission lines in individual sources}
\label{sec:em-indiv}

The equivalent widths of the \oii\ line in the individual spectra were
measured using the \textsc{specfit} package implemented in
\textsc{iraf} \citep{kri94}.  We normalised the spectra by a spline
fit to the continuum, then measured the equivalent width of the
emission line by fitting a Gaussian profile to the line. We must be
careful to account for sources where the emission line fell on a sky
line, or was redshifted out of the spectrum, and so would not have
been detected if it were present. There were 333 sources in which
\oii\ was potentially detectable, with sources at redshifts $\sim0.48$
and $\sim0.58$ most strongly affected when the \oii\ line fell on a
bright sky line.

Following \citet{scm+07}, we used a rest equivalent width cutoff of
7\AA\ to define sources with \oii\ emission.  Using this criterion, 50
out of the 333 (15\%) radio-loud sources in which \oii\ emission was
potentially detectable show such emission, compared to 25 out of the
matched radio-quiet sample (7.5\%); so the radio-loud sources are
significantly more likely to show \oii\ emission than galaxies matched
in optical luminosity and redshift. If we assume the fraction of
radio-quiet galaxies showing emission represents the null-hypothesis
(25/333), then observing 50/333 radio-loud sources with emission is
highly significant ($p \ll 0.001$).  For the \neiii~3869 and \nev~3426
lines, we used a cut-off of $W_\lambda>4$\AA\ to define the
emission-line sources. There were 25 sources with \neiii\ emission and
15 with \nev\ emission in the radio-loud sample, compared to 11 and 8
respectively in the radio-quiet sample. Again, the radio-loud sources
are roughly twice as likely to show emission than their radio-quiet
counterparts (probability $p<0.001$ and 0.05 respectively). Twenty-one
radio-loud sources have $W_\lambda(\oiii)>15$\AA, compared with just
five sources among the radio-quiet galaxies.

\begin{figure}
     \centerline{\psfig{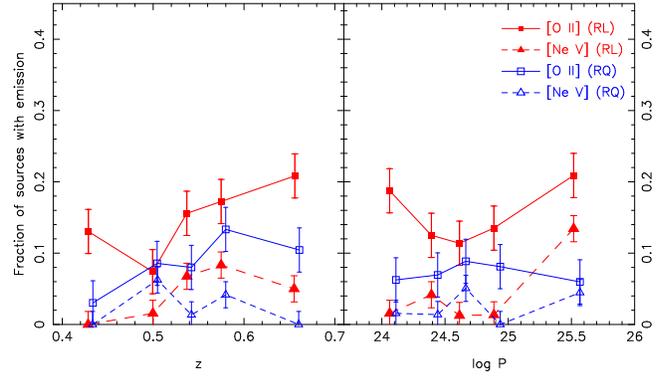}}
     \caption{Fraction of sources with \oii\ and \nev\ emission, as a
       function of redshift and radio power; $\log P$ is defined for
       the radio-quiet sample by the matching to the radio-loud
       sources (\S~\ref{sec:matched}). The red filled symbols
       represent the radio-loud sources (RL), while the blue open
       symbols represent the matched sample of radio-quiet sources
       (RQ). The squares and solid lines show the fraction of sources
       with EW(\oii~3727)$>7$\AA, the triangles and dashed lines show
       the fraction of sources with EW(\nev~3426)$>4$\AA. The
       radio-loud sources show a higher fraction of sources with
       emission than the radio-quiet sources, as well as a tendency
       for this fraction to increase with redshift. The likelihood of
       showing emission increases with radio-power (right-hand panel),
       especially for the \nev\ emission. }
     \label{fig:emfrac}
\end{figure}

There is a slight trend for the fraction of sources showing emission
to increase with redshift. In Fig.~\ref{fig:emfrac} we plot the
fraction of sources which show \oii\ emission (out of those in which
it was potentially detectable) as a function of redshift. A similar
trend is seen in the radio-quiet sources. This could be an absolute
magnitude effect, as we are selecting slightly more luminous galaxies
at higher redshift \citep{scm+07}.

\oii\ emission is seen in radio galaxies at all radio powers (though
it is somewhat more common in low- ($\log P < 24.5$) and high-power
($\log P > 25$) sources, as shown in the right-hand panel of
Fig.~\ref{fig:emfrac}). ``Strong'' \oii\ emission, however, with
$W_\lambda>15$\AA, is significantly more common in the highest-power
radio galaxies: 28\% of sources with $\log P > 25.5$\ (9 out of 35)
have $W_\lambda>15$\AA\ compared to $\sim3$\% of sources at lower
radio power.  \neiii\ and \nev\ emission is also more common at higher
radio power.  Neither of these trends can be seen in the sample of
radio-quiet sources which match the distribution of the radio galaxies
in $z$ and $r$.


\subsubsection{Emission lines in the composite spectra}
\label{sec:em-comp}

By combining many spectra, we can measure changes in line ratios much
more accurately than we can in individual sources.

\begin{figure}
     \centerline{\psfig{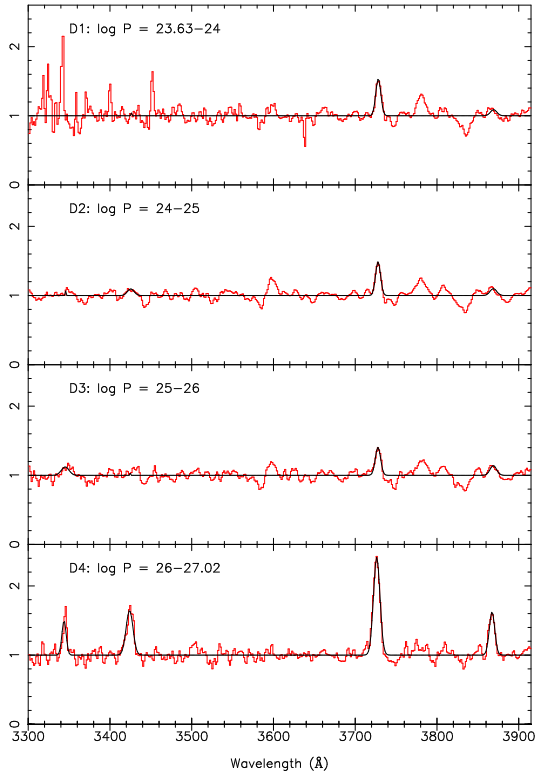}}
     \caption{Gaussian fits to the emission lines in the D composites,
       with equal spacing in radio power. The normalised composite
       spectra are shown in red, and the Gaussian fits to the emission
       lines are shown as the thick black line. \oii\ emission is seen
       at all radio powers, though it is significantly stronger in the
       highest power sources; these are the only ones to show
       significant \nev\ and \neiii\ emission. The fourth line is a
       weaker \nev\ line at 3346\AA. The blue end of the lowest-power
       composite is noisy, because the lowest-power sources are only
       found at nearby redshifts, so there is not much signal at short
       wavelengths in this composite. }
     \label{fig:em-comp}
\end{figure}

We measured the line properties of the emission lines in the composite
spectra by fitting Gaussian profiles using \textsc{specfit} again,
using the same technique as in \S\ref{sec:em-indiv}.  The spectra were
normalised by a spline fit to the continuum, and then Gaussians were
fit, constraining the wavelengths to be within $\pm 5$\AA\ of the
known wavelengths of the lines. These fits are shown in
Figure~\ref{fig:em-comp}; the line properties are listed in
Table~\ref{tab:emline-fits}.

\begin{table}
     \caption{Equivalent widths of the fits to the emission lines in
     the D composites (Fig.~\protect{\ref{fig:em-comp}}) The equivalent
     widths of the three emission lines are shown, for the radio-loud
     and matched radio-quiet composites. The last two lines show the
     fits to the Neon composite and the [O II] composite
     (\S\ref{sec:em-comp}). }
     \label{tab:emline-fits}
\begin{tabular}{cr@{\,$\pm$\,}l r@{\,$\pm$\,}l r@{\,$\pm$\,}l}
\hline
          & \multicolumn{2}{c}{\oii\ 3727} &
            \multicolumn{2}{c}{\nev\ 3426} & 
            \multicolumn{2}{c}{\neiii\ 3869} \\
Composite & \multicolumn{2}{c}{(\AA)}    & 
            \multicolumn{2}{c}{(\AA)}    & 
            \multicolumn{2}{c}{(\AA)}      \\
\hline
Radio loud D1 &  4.2 & 0.7  &  \multicolumn{2}{l}{$< 1.0$}  &  0.9 & 0.7 \\
Radio loud D2 &  3.7 & 0.4  &  1.0 & 0.4  &  1.0 & 0.4 \\
Radio loud D3 &  3.3 & 0.4  &  \multicolumn{2}{l}{$< 0.5$}  &  1.4 & 0.4 \\
Radio loud D4 & 14.1 & 0.6  &  6.8 & 0.5  &  5.2 & 0.4 \\
\\
Radio quiet D1 & 1.5 & 0.4  &   \multicolumn{2}{l}{$< 0.5$} & \multicolumn{2}{l}{$< 0.6$} \\
Radio quiet D2 & 2.1 & 0.3  &   \multicolumn{2}{l}{$< 0.4$} & 1.2 & 0.4 \\
Radio quiet D3 & 2.2 & 0.3  &   \multicolumn{2}{l}{$< 0.3$} & 1.5 & 0.4 \\
Radio quiet D4 & 1.8 & 0.5  &   1.9 & 0.9  &   1.0 & 0.9 \\
\\
Neon composite  & 17.2 & 0.7  &  7.9 & 0.6  &  8.1 & 0.6 \\
\oii\ composite & 17.7 & 0.9  &  \multicolumn{2}{l}{$< 0.7$} & 1.8 & 0.6 \\
\hline
\end{tabular}
\end{table}

\begin{figure}
     \centerline{\psfig{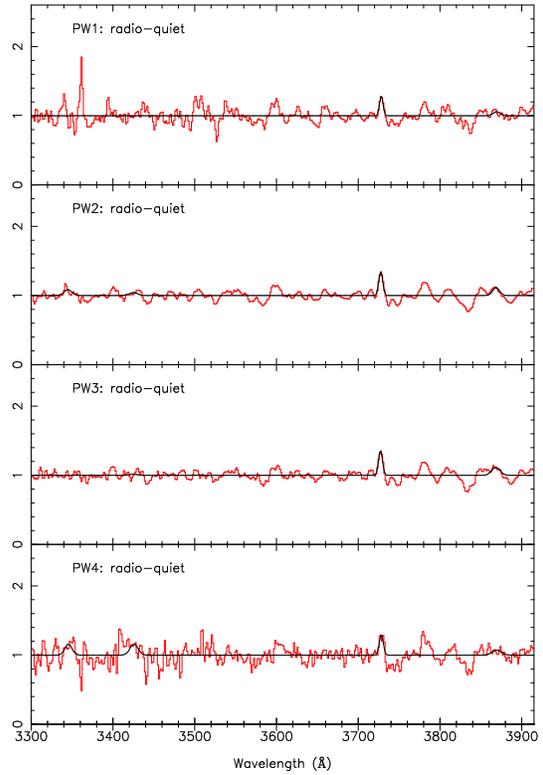}}
     \caption{Same as Fig.~\ref{fig:em-comp}, but for the matched
     radio-quiet composites. Clearly the emission line properties are
     very different to the radio-loud sources, with no detectable
     \nev\ or \neiii\ emission.}
     \label{fig:em-comp-RQ}
\end{figure}

There is little difference in the emission lines of the composite
spectra until the very highest radio-power is reached.  This composite
shows much stronger \oii\ emission, as well as strong \nev\ and
\neiii\ emission lines, lines which are characteristic of quasars. The
radio-quiet galaxies which are matched in luminosity and redshift show
no increase in \oii\ emission, and no trace of neon lines. This is not
just due to the contribution from a single source: three of the nine
sources with highest radio power have $W_{3426}>4$\AA\ (in two sources
the line fell on a sky line).  This result implies that the increased
emission seen in the most powerful radio sources is not just a
function of galaxy size, but is tied to the activity of the central
black hole.

\citet{olk95} and \citet{bkh+05}, found no correlation between radio
power and emission line properties; their samples only reached radio
powers $P \sim 10^{25}\;\mathrm{W/Hz}$. The emission-line luminosity
of very powerful radio sources is known to correlate with radio power
\citep[e.g.][Fig. 2]{mcc93}. There must be a threshold radio power at
which strong emission lines ``switch on'' at a level above the usual
incidence in quiescent galaxies; at the highest radio luminosities,
virtually every radio galaxy has strong forbidden lines \citep{hl79}.
Our results suggest we have found this threshold, at radio powers of
around $10^{26}\;\mathrm{W/Hz}$.

\begin{figure}
     \centerline{\psfig{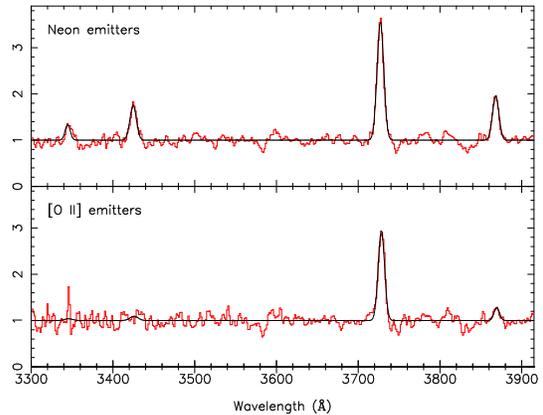}}
     \caption{Gaussian fits to the emission lines in composites of
       sources which show significant \nev\ and \neiii\ emission, and
       sources which show strong \oii\ emission but no \nev\ or
       \neiii\ emission. Only sources where all emission lines were
       clear of the night sky lines were included.  }
     \label{fig:NeO-em}
\end{figure}

As noted earlier (\S~\ref{sec:em-indiv}), there were 21 sources which
showed strong \oii\ emission, with $W_\lambda>15$\AA. There were also
32 sources which showed \neiii\ and/or \nev\ emission. There was not
total overlap between these: there were ten sources which showed
strong \oii\ emission but no \neiii/\nev\ emission. To test whether
this difference was real, we constructed two composite spectra:
\emph{(a)} the 32 sources which showed \nev\ and/or \neiii\ emission
(the ``Neon composite''), and \emph{(b)} the sources that showed
strong \oii\ emission but no \neiii\ or \nev\ emission (the ``\oii\
composite''). We could only use five of the ten sources without
\neiii/\nev\ emission in constructing the \oii\ composite, because the
other five had at least one of the three neon lines falling on a sky
line. The composites are shown in Figure~\ref{fig:NeO-em}.

Clearly these two composites have significantly different line ratios:
the non-detection of \nev\ in some sources is not just a
signal-to-noise ratio issue. The Neon composite has line ratios
$\mathrm{\nev/\oii} = 0.46 \pm 0.03$ and $\mathrm{\neiii/\oii} = 0.47
\pm 0.03$, while the \oii\ composite has $\mathrm{\nev/\oii} < 0.04$\ 
and $\mathrm{\neiii/\oii} = 0.10 \pm 0.03$. This suggests different
ionisation mechanisms are operating in these two groups. For
comparison, the composite quasar spectrum published by \citet{vrb+01}
has ratios $\mathrm{\nev/\oii} = 0.78 \pm 0.02$\ and
$\mathrm{\neiii/\oii} = 0.88 \pm 0.02$.

We investigated whether the stellar continuum in these emission line
composites were at all different; nothing definitive could be
found. The \oii\ composite required a higher proportion of young stars
to model the continuum than the Neon composite (80\% compared to
40\%), but this was only marginally significant; there were only five
spectra in the \oii\ composite.

\section{Conclusions}
\label{sec:conclusions}

We have investigated the optical properties of a volume-limited sample
of radio galaxies in the 2SLAQ Luminous Red Galaxy survey. We created
composite spectra in order to investigate how the stellar populations
of these galaxies depend on redshift or radio power. By constructing a
comparison sample of galaxies matched in redshift and luminosity, we
can eliminate the possibility that any effects we find are a result of
the fact that radio-loud galaxies are on the average more luminous
than radio-quiet galaxies. 

We modelled the underlying starlight using single-age stellar
population models. None of the spectra were adequately described by a
single model; they all demanded an additional stellar component. We
achieved satisfactory fits assuming that each spectrum consisted of an
``old'' (7~Gyr) population plus a fraction of light contributed by a
``young'' population, whose age was allowed to vary between 10~Myr and
5~Gyr. Neither a power-law component nor a model with continuous star
formation described the spectra as well as the two-stellar population
model.

The stellar populations of all the composites were almost identical,
independent of redshift, radio power, and whether or not the galaxy
contains an active radio source.  The fraction of light contributed by
the young population ($\sim$ 40\% at 4050\AA) increased marginally
with redshift, but is not correlated with radio power. The age of the
``young'' component was consistent with an age of
$1000^{+1000}_{-500}$~Myr for all composites. The sole exception was
the composite made from the eleven most powerful radio sources, with
$\log P > 26$, which was fitted with a young population of age
100~Myr. This was not seen in the corresponding composite of
radio-quiet sources, which means it is unlikely to be due to a
systematic effect, which would affect both radio-loud and radio-quiet
galaxies equally. In addition, \citet{scm+07} showed that the median
redshift of 2SLAQ radio galaxies is independent of radio power, so
there are unlikely to be subtle aperture effects affecting the sample.

The two-population model is probably not a very realistic description
of the underlying stellar light. However, our study is a
\emph{differential} test, comparing the properties of the radio-loud
galaxies with their radio-quiet counterparts. In a future paper, we
will investigate other models to represent the stellar populations of
these galaxies.

The emission line properties of the population show more correlation
with radio activity. Approximately 15\% of the radio-loud sources show
significant \oii\ emission, compared to 7.5\% of the radio-quiet
sample; similarly, the radio-loud sources were roughly twice as likely
to show \neiii\ and/or \nev\ emission than their radio-quiet
counterparts (12\% and 6\%, respectively). Still, it is notable that
most of the radio galaxies in the 2SLAQ would not have been recognised
as AGN on the basis of their optical spectra alone. The different
\oii/Ne line ratios seen in different objects suggest there is more
than one emission mechanism operating to ionise the gas: these are
likely to include hot stars and post-AGB stars as well as ionisation
by the central AGN.

These observations support the model that all luminous red galaxies
contain a central black hole, only some of which are active at any one
time; the current activity of the central black hole, as measured by
the radio emission, does not appear to correlate with the stellar
population in the galaxy. The sole exception appears to be that the
most powerful radio galaxies show more activity and younger stars than
the rest of the population. Even so, with young ($\sim
100~\mathrm{Myr}$) stars contributing 30\% of the light, the fraction
of mass in young stars is only 0.5\% of the mass of the galaxy.

We need to be careful to distinguish between differences related to
the activity of the central black hole and luminosity effects, since
more massive galaxies are more likely to harbour AGN. We controlled
for this bias by constructing a matched sample of radio-quiet
galaxies.  The comparison sample is matched in redshift and apparent
magnitude, so the radio-quiet galaxies match the radio-loud sources
closely in absolute luminosity. Since we saw differences in the most
powerful radio galaxies, but not in their radio-quiet counterparts,
the difference most likely relates to the presence of an active black
hole. In Paper~I, the 2SLAQ radio galaxies were found to cluster more
strongly than the 2SLAQ LRG population as a whole \citep{scm+07}. It
was speculated in that paper that this might be a luminosity effect,
since the 2SLAQ radio galaxies are more luminous than the rest of the
sample. However, recent work using samples matched in luminosity
suggest this is not the explanation (Wake et~al., in prep.). This
suggests the radio galaxies really are preferentially located in
denser environments, with more gas available to feed the black hole.
\citet{blk+07} have recently suggested that a similar effect is seen
in lower-redshift radio galaxies from SDSS.

Our results support the findings of \citet{bkh+05}, who found that the
probability that a galaxy is radio-loud is independent of whether it
is optically classified as an AGN, and that the host galaxies of
radio-loud AGN match their radio-quiet counterparts. If optical
emission lines trace current feeding of black holes, while radio
emission points to mostly dormant black holes, accreting at low rates,
then the current rate of growth of black holes in these galaxies is
very low.

One notable point is that the age of the ``young'' population remains
unchanged at 700~Myr (Table~\ref{tab:composites}) despite the
different ages of the galaxies: the highest redshift galaxies in our
sample ($z=0.76$) are 3.1~Gyr younger than the lowest redshift
galaxies ($z=0.31$).  If the stellar populations in the lowest
redshift galaxies were born 1~Gyr before $z=0.31$, we should see the
era of star birth, since it is covered in our sample (at a redshift
$z=0.39$).  We see no evidence, however, for an epoch of star
formation at that redshift.

So where are the galaxies which are forming stars as we observe them?
There are two possibilities. If the single-age stellar population is
a bad assumption, we may have low-level ongoing star-formation instead
of star formation in discrete bursts. The second possibility, however,
is that the initial colour selection of the 2SLAQ sample is biased
towards accepting galaxies with these ages; in other words, the
counterparts of our $z=0.3$ galaxies are indeed forming stars at
$z=0.4$, but they don't appear as red galaxies while forming stars, so
are not included in the 2SLAQ sample.

In fact, our modelling suggests that we may not notice these galaxies
using the 2SLAQ \emph{gri} colour selection criteria. A model galaxy
with 5\% of the mass in young stars formed in a single burst (which
contributes 40--50\% of the light around 1~Gyr after the burst) is
only 1 mag bluer than the bulk of old galaxies. Furthermore, since
most of the colour evolution occurs very early on, in the first
50~Myr, the chances of finding such an object in our sample is not
high. In order to locate these galaxies, we would need to assemble a
large sample of luminous galaxies using a different colour selection.

\section*{Acknowledgments}
We would like to thank the referee, Jasper Wall, for his extremely
helpful comments and suggestions. 
Funding for the SDSS and SDSS-II has been provided by the Alfred P.
Sloan Foundation, the Participating Institutions, the National Science
Foundation, the U.S. Department of Energy, the National Aeronautics
and Space Administration, the Japanese Monbukagakusho, the Max Planck
Society, and the Higher Education Funding Council for England.  The
SDSS Web site \hbox{is \texttt{http://www.sdss.org/}}. This work was
partially supported by National Science Foundation grant AST-0607634 
(N.P.R.).


\end{document}